\documentclass[showpacs,preprintnumbers,10pt,twocolumn]{revtex4}%
\usepackage{amssymb}
\usepackage{amsfonts}
\usepackage{amsmath}
\usepackage{graphicx}
\usepackage{dcolumn}
\usepackage{soul}
\usepackage{bm}
\usepackage{revsymb}
\usepackage{xcolor}%
\begin{document}
\title{ Energy-based theory of autoresonance in chains of coupled damped-driven
generic oscillators}
\author{Ricardo Chac\'{o}n$^{1}$, Faustino Palmero$^{2}$, Pedro J. Mart\'{\i}nez$^{3}%
$, and Somnath Roy$^{4}$}
\affiliation{$^{1}$Departamento de F\'{\i}sica Aplicada, E.I.I., Universidad de
Extremadura, Apartado Postal 382, E-06006 Badajoz, Spain and Instituto de
Computaci\'{o}n Cient\'{\i}fica Avanzada (ICCAEx), Universidad de Extremadura,
E-06006 Badajoz, Spain}
\affiliation{$^{2}$Grupo de F\'{\i}sica No Lineal, Departamento de F\'{\i}sica Aplicada I,
Escuela T\'{e}cnica Superior de Ingenier\'{\i}a Inform\'{a}tica, Universidad
de Sevilla, Avda Reina Mercedes s/n, E-41012 Sevilla, Spain}
\affiliation{$^{3}$Departamento de F\'{\i}sica Aplicada, E.I.N.A., Universidad de Zaragoza,
50018 Zaragoza, Spain}
\affiliation{$^{4}$Department of Physics, Jadavpur University, West Bengal, Kolkata-700032, India}
\date{\today}

\begin{abstract}
An energy-based theory of autoresonance in driven dissipative chains of
coupled generic oscillators is discussed on the basis of a variational
principle concerning the energy functional. The theory is applied to chains of
delayed Duffing-Ueda oscillators and the equations that together govern the
autoresonance forces and solutions are derived and solved analytically for
generic values of parameters and initial conditions, including the case of
quenched time-delay disorder. Remarkably, the presence of retarded potentials
with time-delayed feedback drastically modify the autoresonance scenario
preventing the growth of the energy oscillation over specific regions of the
parameter space. Additionally, effective harmonic forces with a slowly varying
frequency are derived from the exact autoresonant excitations and the
effectiveness of the theory is demonstrated at suppressing the chaos induced
by homogeneous periodic excitations in such oscillator chains. Numerical
experiments confirmed all the theoretical predictions.

\end{abstract}
\keywords{Nonlinear dynamics and nonlinear dynamical systems. Nonlinear resonance.}
\pacs{05.45. -a}
\maketitle

\textit{Introduction.}$-$ Autoresonance (AR) induced secular growth of the
oscillation energy in nonlinear damped-driven systems take place when the
system persistently adjusts its amplitude so that its instantaneous nonlinear
period balances the driving period. Firstly studied in Hamiltonian systems, AR
phenomena have been investigated since the middle of the last century and have
been noted in diverse contexts: particle accelerators \cite{1,2}, atomic and
molecular physics \cite{3}, planetary dynamics \cite{4}, nonlinear
low-dimensional oscillators \cite{5}, nanoscale magnetic devices \cite{6},
nonlinear chains subject to localized driving \cite{7}, plasma waves \cite{8},
and energy transfer between axions \cite{9}, to quote a few instances. While
most previous investigations on AR have restricted themselves to explore the
effectiveness of chirped harmonic forces \cite{1,2,3,4,5,6,7,8,9,10}, a
general energy-based AR (EBAR) theory has been proposed to explain the
approximate and phenomenological findings arising from a prior adiabatic
approach to AR in Duffing-like oscillators \cite{11}. Further application of
the EBAR theory to low-dimensional systems has included the case where the
system crosses a separatrix associated with its underlying integrable
counterpart \cite{12} as well as the problem of chaotic escape in dissipative
multistable systems \cite{13}. Since the application of chirped harmonic
forces to give rise to reliable autoresonant responses in multi-particle
chains seems to be more problematic than in the case of an isolated oscillator
\cite{14}, the question naturally arises: How does an extension of the EBAR
theory to driven dissipative chains of coupled generic oscillators work?

\textit{Theory.}$-$In this Letter, this fundamental problem is studied in the
context of the family of $N$ linearly coupled, identical oscillators
\begin{equation}
\overset{..}{u}_{n}+V^{\prime}(u_{n})=-\eta\overset{.}{u}_{n}+\lambda
\Delta_{d}u_{n}+F_{n}(t), \label{eq1}%
\end{equation}
where dots indicate time derivative, $\Delta_{d}u_{n}\equiv u_{n+1}%
+u_{n-1}-2u_{n}$ is the discrete Laplacian operator, $V^{\prime}(u_{n})\equiv
dV/du_{n}$ with $V(u_{n})$ being a general on-site potential, $\eta$ and
$\lambda$ are the damping coefficient and coupling constant, respectively,
while $F_{n}(t)$ is a temporal force. Clearly, the corresponding equation for
the energy is%
\begin{align}
\overset{.}{E}  &  =\sum_{n}\overset{.}{u}_{n}\left[  -\eta\overset{.}{u}%
_{n}+F_{n}(t)+\lambda\Delta_{d}u_{n}\right] \nonumber\\
&  +\lambda\sum_{n}\left(  \overset{.}{u}_{n+1}-\ \overset{.}{u}_{n}\right)
\left(  u_{n+1}-u_{n}\right) \nonumber\\
&  \equiv P\left(  u_{1},...,u_{N},\overset{.}{u}_{1},...,\overset{.}{u}%
_{N},t\right)  , \label{eq2}%
\end{align}
where $E\equiv\sum_{n}\left[  \overset{.}{u}_{n}^{2}/2+V\left(  u_{n}\right)
\right]  +\frac{\lambda}{2}\sum_{n}\left(  u_{n+1}-u_{n}\right)  ^{2}$ and
$P\left(  u_{1},...,u_{N},\overset{.}{u}_{1},...,\overset{.}{u}_{N},t\right)
$ are the energy and power, respectively. As in the case of isolated
oscillators, the AR solutions of Eq. (1) are defined in terms of a variational
principle by imposing that the energy variation $\Delta E=\int_{t_{1}}^{t_{2}%
}P\left(  u_{1},...,u_{N},\overset{.}{u}_{1},...,\overset{.}{u}_{N},t\right)
dt$ is a maximum (with $t_{1},t_{2}$ arbitrary but fixed instants) under the
influence of dissipation and forcing. Thus, the corresponding Euler-Lagrange
equations provide the necessary conditions (AR conditions) to be fulfilled by
the AR solutions and temporal forces:
\begin{equation}
\frac{\partial P}{\partial u_{n}}-\frac{d}{dt}\left(  \frac{\partial
P}{\partial\overset{.}{u}_{n}}\right)  =0, \label{eq3}%
\end{equation}
$n=1,...,N$. Indeed, Eq. (\ref{eq3}) supply relationships between $u_{n}$,
$\overset{.}{u}_{n}$, and $F_{n}$ such that the solutions of the system given
by Eqs. (\ref{eq1}) and (\ref{eq3}) together provide the AR forces,
$F_{n,AR}(t)$, and the AR solutions, $u_{n,AR}\left(  t\right)  $, for each
set of initial conditions:%
\begin{align}
\overset{..}{u}_{n,AR}+V^{\prime}(u_{n,AR})  &  =\eta\overset{.}{u}%
_{n,AR}+\lambda\Delta_{d}u_{n,AR},\label{eq4}\\
F_{n,AR}(t)  &  =2\eta\overset{.}{u}_{n,AR}, \label{eq5}%
\end{align}
$n=1,...,N$. Clearly, this AR scenario is strongly dependent upon the
distribution of initial conditions: while for non-uniform distributions the AR
forces and solutions depend upon the coupling constant through the discrete
Laplacian operator, for uniform distributions the coupling energy is always
zero (a situation equivalent to that of the anticontinous limit $\lambda=0$),
and hence the AR forces and solutions are those corresponding to the
respective isolated oscillators. Regarding the Hamiltonian limiting case
$\eta\rightarrow0$, notice that Eq. (\ref{eq4}) can be equivalently rewritten
as $\overset{..}{u}_{n,AR}+V^{\prime}(u_{n,AR})=F_{n,AR}/2+\lambda\Delta
_{d}u_{n,AR}$ [cf. Eq. (\ref{eq5})], which suggests the natural ansatz
$F_{n,AR}(t)=\kappa\overset{.}{u}_{n,AR},\kappa>0$, for Hamiltonian chains
$\left(  \eta=0\right)  $ and where the AR rate, $\kappa$, is a free parameter
controlling the initial force strength. Therefore, one can expect that the AR
solutions and forces for the Hamiltonian case are essentially the same than
those for the dissipative case, both with $\kappa$ instead of $\eta$.

\begin{figure}[h]
\centering
\includegraphics[width=0.45\columnwidth]{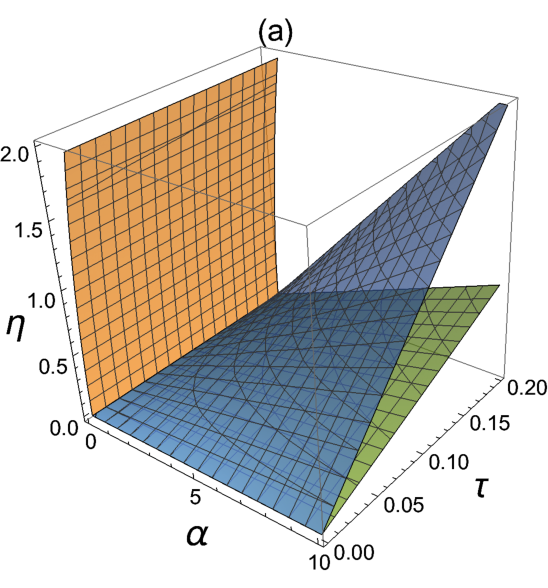}
\includegraphics[width=0.45\columnwidth]{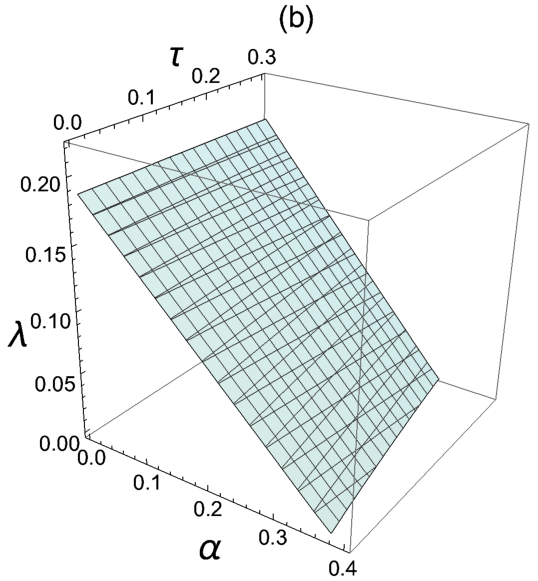}\caption{(a) Constrain
condition for $\lambda=0$ [Eq. (\ref{eq10})] (leftmost surface), critical
damping $\eta=\eta_{c}\equiv\alpha\tau$, and critical damping $\eta
=\eta_{c,mf}\equiv\sqrt{\alpha/2}\ln\left(  1+\alpha\tau_{\max}^{2}/2\right)
/\arctan\left(  \sqrt{\alpha/2}\tau_{\max}\right)  $ (rightmost surface). (b)
Constrain condition for $\eta=1.3$ [Eq. (\ref{eq10})]. }%
\label{fig1}%
\end{figure}

\textit{Duffing-Ueda oscillators with time-delayed feedback.}$-$ Let us
consider the application of the above EBAR theory to the significant instance
of purely nonlinear (cubic) oscillators with a homogeneous retarded potential
with time-delayed feedback $V^{\prime}(u_{n})=\beta u_{n}^{3}+\alpha
u_{n}\left(  t-\tau\right)  $ [cf. Eq. (\ref{eq1})], with the positive
parameters $\alpha$ and $\tau$ accounting for the strength and time-delay of
the retardation term, respectively. Time delays are unavoidable in real-world
systems since they are induced because of the finite time needed to exchange
information in complex (coupled) systems \cite{15}. After assuming that $\tau$
is sufficiently small such that $u_{n}\left(  t-\tau\right)  =u_{n}%
(t)-\tau\overset{.}{u}_{n}\left(  t\right)  +\tau^{2}\overset{..}{u}%
_{n}\left(  t\right)  /2+O\left(  \tau^{3}\right)  $, Eqs. (\ref{eq4}) and
(\ref{eq5}) become%
\begin{align}
\overset{..}{u}_{n,AR}+\omega_{0}^{2}u_{n,AR}+Bu_{n,AR}^{3}  &  =\delta
\overset{.}{u}_{n,AR}+\Lambda\Delta_{d}u_{n,AR},\label{eq6}\\
F_{n,AR}(t)  &  =2\delta R\overset{.}{u}_{n,AR}, \label{eq7}%
\end{align}
$n=1,...,N$, with $\omega_{0}^{2}\equiv\alpha/R,R\equiv1+\alpha\tau
^{2}/2,B\equiv\beta/R,\delta\equiv\left(  \eta-\alpha\tau\right)
/R,\Lambda\equiv\lambda/R$. Thus, when $N$ is a multiple of 4 \cite{16} and
periodic boundary conditions are assumed, Eqs. (\ref{eq6}) and (\ref{eq7})
present AR solutions and forces
\begin{align}
u_{n,AR}(t)  &  =\gamma_{0}\operatorname{e}^{\delta t/3}\operatorname{cn}%
\left[  \varphi_{n}\left(  t\right)  ;%
%TCIMACRO{\U{bd}}%
%BeginExpansion
\frac12
%EndExpansion
\right]  ,\label{eq8}\\
F_{n,AR}(t)  &  =\frac{2}{3}\gamma_{0}\delta^{2}R\operatorname{e}^{\delta
t/3}\operatorname{cn}\left[  \varphi_{n}\left(  t\right)  ;%
%TCIMACRO{\U{bd}}%
%BeginExpansion
\frac12
%EndExpansion
\right] \nonumber\\
&  -2\gamma_{0}^{2}\delta R\sqrt{B}\operatorname{e}^{2\delta t/3}%
\operatorname{sn}\left[  \varphi_{n}\left(  t\right)  ;%
%TCIMACRO{\U{bd}}%
%BeginExpansion
\frac12
%EndExpansion
\right]  \operatorname*{dn}\left[  \varphi_{n}\left(  t\right)  ;%
%TCIMACRO{\U{bd}}%
%BeginExpansion
\frac12
%EndExpansion
\right]  , \label{eq9}%
\end{align}
with the constraint $\Lambda=\delta^{2}/9-\omega_{0}^{2}/2$, i.e.,
\begin{equation}
\left(  1+\frac{\alpha\tau^{2}}{2}\right)  (\alpha+2\lambda)=\frac{2}{9}%
(\eta-\alpha\tau)^{2} \label{eq10}%
\end{equation}
(see Fig (\ref{fig1})), and where $\operatorname{cn}$, $\operatorname{sn}$,
$\operatorname*{dn}$ are Jacobian elliptic functions of parameter $m$,
$\varphi_{n}\left(  t\right)  \equiv3\gamma_{0}$ $\delta^{-1}B^{1/2}%
\operatorname{e}^{\delta t/3}+n\operatorname{K}(1/2)+\phi_{0}$, with
$\gamma_{0}$, $\phi_{0}$ being arbitrary constants and $\operatorname{K}(m)$
the complete elliptic integral of the first kind, while the initial conditions
satisfy the relationships $u_{n,AR}(0)=\gamma_{0}\operatorname{cn}\left[
\varphi_{n}\left(  0\right)  ;%
%TCIMACRO{\U{bd}}%
%BeginExpansion
\frac12
%EndExpansion
\right]  ,\overset{.}{u}_{n,AR}(0)=\frac{1}{3}\gamma_{0}\delta
R\operatorname{cn}\left[  \varphi_{n}\left(  0\right)  ;%
%TCIMACRO{\U{bd}}%
%BeginExpansion
\frac12
%EndExpansion
\right]  -\gamma_{0}^{2}\sqrt{B}\operatorname{sn}\left[  \varphi_{n}\left(
0\right)  ;%
%TCIMACRO{\U{bd}}%
%BeginExpansion
\frac12
%EndExpansion
\right]  \operatorname*{dn}\left[  \varphi_{n}\left(  0\right)  ;%
%TCIMACRO{\U{bd}}%
%BeginExpansion
\frac12
%EndExpansion
\right]  $. 

\begin{figure}[hh]
\centering
\includegraphics[width=0.95\columnwidth]{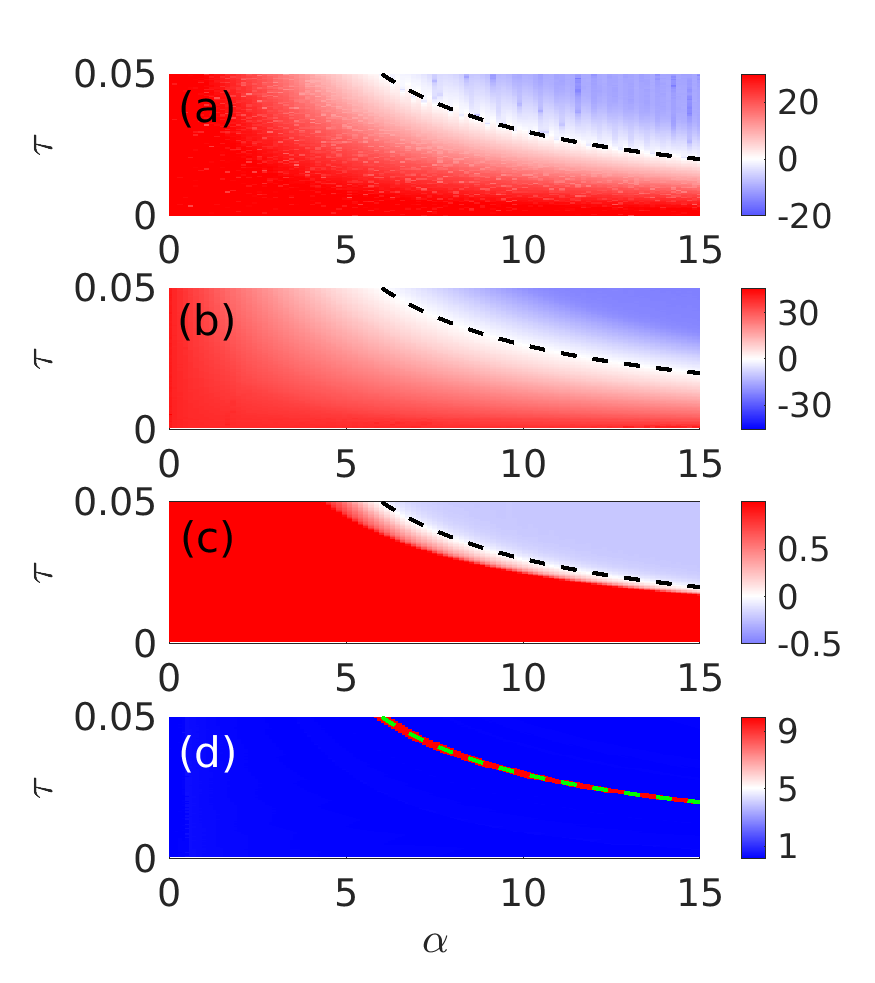} \caption{(a), (b), (c)
Energy amplification strength, $\log\left[  E\left(  t=t_{\max}\right)
/E\left(  t=0\right)  \right]  $, in the $\alpha-\tau$ parameter plane for a
ring of $N=15$ oscillators with $\beta=1,\eta=0.3$ and $t_{\max}=100$. (a)
Isolated oscillator $\left(  \lambda=0\right)  $. (b) Case where the exact AR
force is applied to all oscillators and $\lambda=0.1$. (c) Case where the
exact AR force is applied to a single oscillator and $\lambda=0.1$. (d)
Dimensionless onset time for AR ($\delta t$), after which the adiabatic
invariant [Eq. (\ref{eq14})] undergoes a deviation equal to $10\%$ with
respect to its initial value. The initial conditions are randomly and
independently chosen. The dashed lines (green (grey) in version (d)) indicate
the critical damping $\eta=\eta_{c}\equiv\alpha\tau$ separating the
amplification regime from the extinction regime.}%
\label{fig2}%
\end{figure}

In the two cases of homogeneous initial conditions and
anticontinous limit, the system [Eq. (\ref{eq6})] reduces to a set of
uncoupled oscillators and the corresponding general AR solutions and forces
are those given by Eqs. (8) and (9) with $\varphi_{n}\left(  t\right)
\equiv3\gamma_{0}$ $\delta^{-1}B^{1/2}\operatorname{e}^{\delta t/3}+\phi_{0}$
for any $N$ and arbitrary boundary conditions, while the constraint reduces to
$\omega_{0}^{2}=2\delta^{2}/9$ \cite{11}. Mathematically, the constraint is
precisely the same condition for Eq. (\ref{eq6}) to present, in the
anticontinous limit, both the Painlev\'{e} property and a nontrivial Lie
symmetry \cite{17}, indicating thus that such an equation is integrable.
Remarkably, there exists a \textit{critical} value of the damping coefficient,
$\eta=\eta_{c}\equiv\alpha\tau$, such that for $\eta>$ $\eta_{c}$ one has
genuine AR solutions and forces given by Eqs. (\ref{eq8}) and (\ref{eq9}),
while for $\eta<$ $\eta_{c}$ the Euler-Lagrange equations [Eq. (\ref{eq3})]
provide a necessary condition for the energy functional to present a minimum
(i.e., that energy corresponding to equilibria). Physically, the constrain
establishes a necessary condition to be satisfied by the strengths of
dissipation, coupling, and retardation parameters for the energy amplification
(decrease) rate to be maximum when $\eta>$ $\eta_{c}$ $\left(  \eta<\eta
_{c}\right)  $. Furthermore, Eq. (\ref{eq6}) can be derived from a Lagrangian
\begin{equation}
\mathcal{L}=\frac{\operatorname{e}^{-\delta t}}{2}\sum_{n}\left[  \overset
{.}{u}_{n}^{2}-\omega_{0}^{2}u_{n}^{2}-\frac{Bu_{n}^{4}}{2}-\Lambda
(u_{n+1}-u_{n})^{2}\right]  , \label{eq11}%
\end{equation}
whose associated Hamiltonian is
\begin{equation}
H=\sum_{n}\left\{  \frac{\operatorname{e}^{\delta t}p_{n}^{2}}{2}%
+\frac{\operatorname{e}^{-\delta t}}{2}\left[  \omega_{0}^{2}u_{n}^{2}%
+\frac{Bu_{n}^{4}}{2}+\Lambda(u_{n+1}-u_{n})^{2}\right]  \right\}  ,
\label{eq12}%
\end{equation}
where $p_{n}\equiv\partial\mathcal{L}/\partial\overset{.}{u}_{n}%
=\operatorname{e}^{-\delta t}\overset{.}{u}_{n}$. For the critical damping
$\eta=\eta_{c}$, the Hamiltonian is time-independent and energy is thereby
conserved. After using the canonical transformation $U_{n}=u_{n}%
\operatorname{e}^{-\delta t/2},P_{n}=p_{n}\operatorname{e}^{\delta t/2}$,
together with the generating function $F_{2}=\sum_{n}u_{n}P_{n}%
\operatorname{e}^{-\delta t/2}$ \cite{18}, the new Hamiltonian reads
\begin{align}
K  &  =\frac{1}{2}\sum_{n}\left[  P_{n}^{2}+\omega_{0}^{2}U_{n}^{2}+\frac
{B}{2}\operatorname{e}^{\delta t}U_{n}^{4}\right] \nonumber\\
&  +\frac{1}{2}\sum_{n}\left[  \Lambda(U_{n+1}-U_{n})^{2}-\delta P_{n}%
U_{n}\right]  , \label{eq13}%
\end{align}
and one obtains (after expanding $\operatorname{e}^{\delta t}$) that the AR
solutions are associated (in terms of the old canonical variables and
parameters) with the \textit{adiabatic} invariant%
\begin{align}
&  \sum_{n}\left(  \frac{p_{n}^{2}}{2}+\frac{\alpha u_{n}^{2}}{2+\alpha
\tau^{2}}+\frac{\beta u_{n}^{4}}{4+2\alpha\tau^{2}}\right) \nonumber\\
&  +\sum_{n}\left[  \frac{\lambda(u_{n+1}-u_{n})^{2}-\left(  \eta-\alpha
\tau\right)  p_{n}u_{n}}{2+\alpha\tau^{2}}\right]  \label{eq14}%
\end{align}
over the time interval $0\leqslant t\lesssim t_{AI}$, $t_{AI}\sim$
$\delta^{-1}\sim\left[  \left(  1+\alpha\tau^{2}/2\right)  /\left(
\lambda+\alpha/2\right)  \right]  ^{1/2}$ [cf. Eq. (\ref{eq10})], with
$t_{AI}$ being the onset time for AR. Thus, $t_{AI}$ provides a time scale
from which the energy amplification effects are noticeable when $\eta>$
$\eta_{c}$. When $\eta\rightarrow\eta_{c}$, Eq. (\ref{eq10}) cannot be
satisfied, while $t_{AI}\rightarrow\infty$ and hence Eq. (14) provides the
aforementioned invariant (energy) due to $F_{n,AR}(t)\rightarrow0$ [cf. Eq.
(\ref{eq9})]. When $\lambda\rightarrow0$, the adiabatic invariant Eq.
(\ref{eq14}) reduces to a set of $N$ identical adiabatic invariants
corresponding to a set of $N$ uncoupled Duffing-Ueda oscillators. Extensive
numerical simulations confirmed all the features of the present AR scenario
[see Fig. (\ref{fig2})]. 

Next, we derive effective chirped harmonic forces from Eq. (\ref{eq9}) for
$\eta>\eta_{c}$. For $t\lesssim t_{b}$, $t_{b}\sim$ $\delta^{-1}\sim\left[
\left(  1+\alpha\tau^{2}/2\right)  /\left(  \lambda+\alpha/2\right)  \right]
^{1/2}$ [cf. Eq. (\ref{eq10})], with $t_{b}$ being the breaking time for AR
\cite{5}, and, e.g., homogeneous initial conditions near equilibria
($u_{i}(0)\simeq0,\overset{.}{u}_{i}\left(  0\right)  \simeq0,i=1,...,N$), one
straightforwardly obtains%
\begin{align}
F_{n,AR}(t)  &  \simeq\frac{2}{3}\gamma_{0}\left(  \eta-\eta_{c}\right)
^{2}\cos\left[  \Omega\left(  t\right)  t\right]  ,\label{eq15}\\
\Omega\left(  t\right)   &  \equiv\frac{3\gamma_{0}\sqrt{2\alpha\beta}%
}{2\left(  \eta-\eta_{c}\right)  }\left[  1+\frac{3\alpha}{4\left(  \eta
-\eta_{c}\right)  }t\right]  \label{eq16}%
\end{align}
which is of the form $\varepsilon\cos\left(  \Omega_{0}t+\xi t^{2}/2\right)  $
\cite{5}, with $\varepsilon$ and $\xi$ being the amplitude and linear sweep
rate, respectively. Therefore, the EBAR theory predicts the following scale
laws for the respective thresholds for AR: $\varepsilon_{th}\sim\left(
\eta-\alpha\tau\right)  ^{2},\xi_{th}\sim\alpha^{3/2}$ $\left(  \eta
-\alpha\tau\right)  ^{-2}$. Notice that a multiple scale analysis only allows
us to recover the scaling for $\varepsilon_{th}$, while chirped harmonic
forces valid for arbitrary initial conditions can be straightforwardly
calculated from Eq. (9) (see Supplemental Material (SM) \cite{19} for further details).

\textit{Time-delay disorder.}$-$We study the effects of quenched disorder on
the above AR scenario by randomly choosing the time-delays $\tau_{n}$
uniformly from the interval $\left[  0,\tau_{\max}\right]  $. Thus, Eqs.
(\ref{eq6}) and (\ref{eq7}) become a family of randomly AR equations (one for
each sampling of the uniform distribution) with $\omega_{0}^{2},B,\delta
,\Lambda$ being disorder-induced random parameters having averages
$\left\langle \omega_{0}^{2}\right\rangle =\tau_{\max}^{-1}\alpha
\sqrt{2/\alpha}\arctan\left(  \sqrt{\alpha/2}\tau_{\max}\right)  ,\left\langle
B\right\rangle =\tau_{\max}^{-1}\beta\sqrt{2/\beta}\arctan\left(  \sqrt
{\beta/2}\tau_{\max}\right)  ,\left\langle \Lambda\right\rangle =\tau_{\max
}^{-1}\lambda\sqrt{2/\lambda}\arctan\left(  \sqrt{\lambda/2}\tau_{\max
}\right)  $, $\left\langle \delta\right\rangle =\tau_{\max}^{-1}\eta
\sqrt{2/\alpha}\arctan\left(  \sqrt{\alpha/2}\tau_{\max}\right)  -\tau_{\max
}^{-1}\ln\left(  1+\alpha\tau_{\max}^{2}/2\right)  $, and where $\left\langle
\cdot\right\rangle \equiv\tau_{\max}^{-1}\int_{0}^{\tau_{\max}}\left(
\cdot\right)  d\tau_{n}$. Therefore, the effective (mean-field) AR equations
reads%
\begin{align}
\overset{..}{u}_{n,AR}+\left\langle \omega_{0}^{2}\right\rangle u_{n,AR}%
+\left\langle B\right\rangle u_{n,AR}^{3}  &  =\left\langle \delta
\right\rangle \overset{.}{u}_{n,AR}+\left\langle \Lambda\right\rangle
\Delta_{d}u_{n,AR},\label{eq17}\\
\left\langle 1/R\right\rangle F_{n,AR}(t)  &  =2\left\langle \delta
\right\rangle \overset{.}{u}_{n,AR}, \label{eq18}%
\end{align}
whose solutions are given by Eqs. (\ref{eq8}) and (\ref{eq9}) with obvious
substitutions. Remarkably, this mean-field approach also predicts the
existence of a critical value of the damping coefficient, $\eta=\eta
_{c,mf}\equiv\sqrt{\alpha/2}\ln\left(  1+\alpha\tau_{\max}^{2}/2\right)
\left[  \arctan\left(  \sqrt{\alpha/2}\tau_{\max}\right)  \right]  ^{-1}$,
such that for $\eta>$ $\eta_{c,mf}$ one has optimal energy amplification on
average, while for $\eta<$ $\eta_{c,mf}$ the chain's energy tends to a minimum
on average, which indicates the \textit{robustness} of the above AR scenario
against the presence of quenched time-delay disorder. Note that $\eta
_{c,mf}\rightarrow\eta_{c}$ with $\tau=\tau_{\max}/2$ when $\tau_{\max
}\rightarrow0$, as expected [cf Fig. (\ref{fig1}(a)]. We found that quenched
time-delay disorder favors the energy's amplification with respect to the
homogeneous case, as predicted from the above mean field approximation and is
confirmed by numerical experiments (see Fig. 3). \begin{figure}[h]
\centering
\includegraphics[width=0.95\columnwidth]{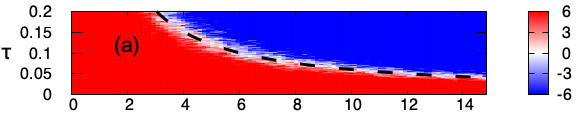}
\includegraphics[width=0.95\columnwidth]{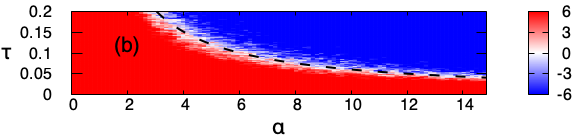} \caption{Energy
amplification strength, $\log\left[  E\left(  t=t_{\max}\right)  /E\left(
t=0\right)  \right]  $, in the $\alpha-\tau_{\max}$ parameter plane for
$\beta=1,\eta=0.3,t_{\max}=50$. (a) Isolated oscillator $\left(
\lambda=0\right)  $. (b) Ring of $N=5$ oscillators with $\lambda=0.1$ and
where the retardation term and the exact AR force are solely applied to a
single oscillator, while the initial conditions are randomly and independently
chosen. The dashed lines indicate the critical damping $\eta=\eta_{c,mf}%
\equiv\sqrt{\alpha/2}\ln\left(  1+\alpha\tau_{\max}^{2}/2\right)  \left[
\arctan\left(  \sqrt{\alpha/2}\tau_{\max}\right)  \right]  ^{-1}$ separating
the amplification regime from the extinction regime.}%
\label{fig3}%
\end{figure}

\begin{figure}[h]
\centering
\includegraphics[width=0.95\columnwidth]{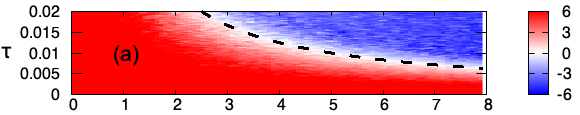}
\includegraphics[width=0.95\columnwidth]{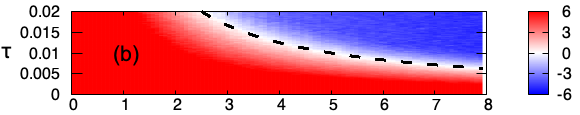}
\includegraphics[width=0.95\columnwidth]{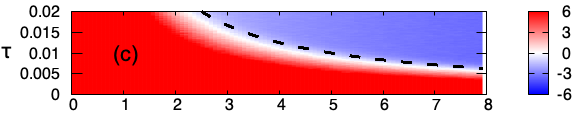}
\includegraphics[width=0.95\columnwidth]{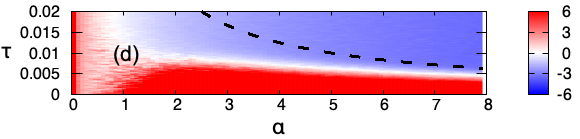} \caption{Energy
amplification strength, $\log\left[  E\left(  t=t_{\max}\right)  /E\left(
t=0\right)  \right]  $, in the $\alpha-\tau$ parameter plane for a ring of
$N=5$ oscillators with $\beta=1,\eta=0.05,\gamma=0.1,\omega=0.21,t_{\max}%
=200$, while the chain presents a chaotic attractor for $\alpha=F_{n}(t)=0$.
(a) Isolated oscillator $\left(  \lambda=0\right)  $. (b) Case where the
retardation term and the exact AR force are applied to all oscillators and
$\lambda=0.1$. (c), (d) Cases where the retardation term and the exact AR
force are solely applied to a single oscillator and (c) $\lambda=0.01$, (d)
$\lambda=0.05$. The dashed lines indicate the critical damping $\eta=\eta
_{c}\equiv\alpha\tau$ separating the amplification regime from the extinction
(regularization) regime (blank region).}%
\label{fig4}%
\end{figure}

\textit{AR-induced chaos suppression.}$-$Since local injection and absorption
of energy can modify a chain%
%TCIMACRO{\U{b4}}%
%BeginExpansion
\'{}%
%EndExpansion
s energy landscapes, and hence reshape the basins of attraction of the
possible attractors in phase space, we study the effectiveness of retardation
terms and AR forces at suppressing the chaos arising from Duffing-Ueda chains
when the oscillators are solely subjected to dissipation and homogeneous
harmonic driving: $\overset{..}{u}_{n}+\beta u_{n}^{3}=-\eta\overset{.}{u}%
_{n}+\gamma\cos\left(  \omega t\right)  +\lambda\Delta_{d}u_{n}$, $n=1,...,N$.
Thus, for the case $V^{\prime}(u_{n})=\beta u_{n}^{3}+\alpha u_{n}\left(
t-\tau\right)  -\gamma\cos\left(  \omega t\right)  $ [cf. Eq. (\ref{eq1})],
Eqs. (\ref{eq4}) and (\ref{eq5}) become
\begin{align}
\overset{..}{u}_{n,AR}+\omega_{0}^{2}u_{n,AR}+Bu_{n,AR}^{3}  &  =\delta
\overset{.}{u}_{n,AR}+\Lambda\Delta_{d}u_{n,AR},\label{eq19}\\
F_{n,AR}(t)  &  =2\delta R\overset{.}{u}_{n,AR}-\gamma\cos\left(  \omega
t\right)  \label{eq20}%
\end{align}
$n=1,...,N$. Clearly, the AR solutions are the same for the cases with and
without homogeneous periodic driving, but the corresponding AR forces differ
precisely in such a periodic force [cf. Eqs. (\ref{eq7}) and (\ref{eq20})].
Now, we explore the effectiveness of locally applying AR forces on $M\ \left(
1\leqslant M\leqslant N\right)  $ oscillators at suppressing the chaos
existing for $\alpha=F_{n}(t)=0$. Two suppressory scenarios are expected
depending on whether $\eta$ is greater or less than $\eta_{c}$. When
$\eta>\eta_{c}$, one has an optimal local injection of energy on each of the
$M$ oscillators subjected to AR forces. Since these $M$ oscillators act as
energy sources for the remaining oscillators, after a certain time interval
$\Delta t=t_{f}-t_{i}$, with $t_{i}$ and $t_{f}$ being the initial and final
instants for the application of the AR forces, and also depending upon the
remaining parameters, the chain's energy $E$ is expected to reach a
sufficiently high value to allow the chain to escape from the basin of the
chaotic attractor (scenario I). However, this scenario is solely expected to
be effective in the presence of multistability, i.e., when the damping
coefficient is sufficiently small to allow the coexistence of a number of
attractors for a fixed set of parameters. On the contrary, one has a
monotonous loss of energy in each of the $M$ oscillators when $\eta<\eta_{c}$,
i.e., they are behaving as energy sinks for the chain's energy. This loss of
energy may again, depending upon the remaining parameters, allow the chain to
escape from the basin of the chaotic attractor, thus regularizing its dynamics
(scenario II). The former scenario is clearly less effective and less
far-reaching than the latter scenario. Figure \ref{fig4} show illustrative
examples confirming the scenario II. 

\textit{Conclusions.}$-$We have developed a general energy-based theory of AR
in damped driven chains of coupled generic oscillators and applied it to
chains of delayed Duffing-Ueda oscillators to reveal a quite complex scenario
of AR. This scenario provides accurate indications on how locally and
optimally control the injection and absorption of energy to modify the global
dynamics of the oscillator chains, in particular, of the order$\leftrightarrow
$chaos transitions, including the case of quenched time-delay disorder. In
contrast to previous approaches to AR in oscillator chains \cite{7}, in which
chirped harmonic forces are systematically used, the present theory provides
exact AR solutions and forces, from which useful chirped harmonic forces are
derived and the limits of their effectiveness in parameter space are
established. Future developments and applications of the present theory
involve the control of topological solitons in Frenkel-Kontorova lattices as
well as the control of dynamics in complex networks of damped-driven nonlinear systems.

\begin{acknowledgments}
Financial support from the Ministerio de Ciencia, Innovaci\'{o}n y
Universidades (MICIU, Spain) through Project No.
PID2019-108508GB-I00/AEI/10.13039/501100011033 cofinanced by FEDER funds
(R.C., F.P.) is gratefully acknowledged.
\end{acknowledgments}


\begin{thebibliography}{99}
\bibitem {16}This comes ultimately from the fact that $4K(m)$ is the real
period of $\operatorname{cn}$.
\end{thebibliography}
\end{document}